\documentclass[conference]{IEEEtran}
\IEEEoverridecommandlockouts
\usepackage{cite}
\usepackage{amsmath,amssymb,amsfonts}
\usepackage{algorithmic}
\usepackage{graphicx}
\usepackage{subfigure}
\usepackage{textcomp}
\usepackage{xcolor}
\usepackage{booktabs}
\usepackage{multirow}

\def\BibTeX{{\rm B\kern-.05em{\sc i\kern-.025em b}\kern-.08em
    T\kern-.1667em\lower.7ex\hbox{E}\kern-.125emX}}

\DeclareRobustCommand*{\IEEEauthorrefmark}[1]{%
    \raisebox{0pt}[0pt][0pt]{\textsuperscript{\footnotesize\ensuremath{#1}}}}

\begin{document}

\title{Estimation of the User Contribution Rate by Leveraging Time Sequence in Pairwise Matching function-point between Users Feedback and App Updating Log}

\author{
\IEEEauthorblockN{
Shiqi Duan\IEEEauthorrefmark{1,2},
Jianxun Liu\IEEEauthorrefmark{1,2},
Yong Xiao\IEEEauthorrefmark{1,2}, and
Xiangping Zhang\IEEEauthorrefmark{1,2}}
\IEEEauthorblockA{\IEEEauthorrefmark{1}School of Computer Science and Engineering, Hunan University of Science and Technology, Hunan, China}
\IEEEauthorblockA{\IEEEauthorrefmark{2}Hunan Key Lab for Services Computing and Novel Software Technology, Hunan, China}

\IEEEauthorblockA{\{duansq379, ljx529, yongx853, hi.xiangping\}@gmail.com}
}

\iffalse
% original author temp
\author{\IEEEauthorblockN{1\textsuperscript{st} Shiqi Duan}
\IEEEauthorblockA{\textit{dept. name of organization (of Aff.)} \\
\textit{name of organization (of Aff.)}\\
City, Country \\
email address or ORCID}
\and
\IEEEauthorblockN{2\textsuperscript{nd} Jianxun Liu}
\IEEEauthorblockA{\textit{dept. name of organization (of Aff.)} \\
\textit{name of organization (of Aff.)}\\
City, Country \\
email address or ORCID}
\and
\IEEEauthorblockN{3\textsuperscript{rd} Yong Xiao}
\IEEEauthorblockA{\textit{dept. name of organization (of Aff.)} \\
\textit{name of organization (of Aff.)}\\
City, Country \\
email address or ORCID}

}
% original author temp
\fi

\maketitle

\begin{abstract}
Mobile applications have become an inseparable part of people's daily life. Nonetheless, the market competition is extremely fierce, and apps lacking recognition among most users are susceptible to market elimination. To this end, developers must swiftly and accurately apprehend the requirements of the wider user base to effectively strategize and promote their apps' orderly and healthy evolution. The rate at which general user requirements are adopted by developers, or user contribution, is a very valuable metric that can be an important tool for app developers or software engineering researchers to measure or gain insight into the evolution of app requirements and predict the evolution of app software. Regrettably, the landscape lacks refined quantitative analysis approaches and tools for this pivotal indicator. To address this problem, this paper exploratively proposes a quantitative analysis approach based on the temporal correlation perception that exists in the app update log and user reviews, which provides a feasible solution for quantitatively obtaining the user contribution. The main idea of this scheme is to consider valid user reviews as user requirements and app update logs as developer responses, and to mine and analyze the pairwise and chronological relationships existing between the two by text computing, thus constructing a feasible approach for quantitatively calculating user contribution. To demonstrate the feasibility of the approach, this paper collects data from four Chinese apps in the App Store in mainland China and one English app in the U.S. region, including 2,178 update logs and 4,236,417 user reviews, and from the results of the experiment, it was found that 16.6\%-43.2\% of the feature of these apps would be related to the drive from the online popular user requirements.
\end{abstract}


\section{Introduction}
With the widespread use of mobile applications (apps) in people's daily lives, the development and maintenance of apps have attracted great attention in the Internet field. To be able to occupy a central position in the mobile Internet industry chain, developers have flocked to the fiercely competitive market of app development. Nonetheless, within the app store, a substantial number of apps offer comparable features, leading to fierce competition. Apps that fail to retain their user base are at risk of being ousted from the market \cite{r1, r2}. Consequently, to maintain a sustained competitive advantage, developers must consistently satisfy user requirements and deliver an exceptional user experience. 

In contrast to the traditional software of the PC era, app store creates a new and open business model that integrates the network and mobile devices. It offers a convenient online interaction platform for both app developers and users. On the one hand, developers can upload apps to app store, including app names, feature profiles, preview screenshots and other information, and then continuously launch new versions that have been perfected and optimized to provide services to users; on the other hand, users can search for and download the corresponding apps according to their requirements, and express their personal views on the app's functions or limitations, expound on their opinions about the app, and give suggestions to the developers. This channel of information exchange not only facilitates the selection of apps by users, but also conveys the ``voice" of the user to the app developer to assist in the release plan \cite{r3, r4}.

Developers leverage user feedback effectively, from which they can extract and refine users' requirements and release versions that make users more satisfied, which helps the development of the app. However, certain apps receive thousands of user reviews a day \cite{r5, r6}, and many reviews contain a mix of developer suggestions and extraneous content, such as user insults and provocations \cite{r5, r7, r8}, and only about one-third of them are useful to developers according to statistical analysis \cite{r5, r6, r9}, and it is obviously impractical to analyze each user review manually. Therefore, existing researches \cite{r5, r8, r10, r11, r12, r13} categorize user reviews based on different application scenarios to extract information that is beneficial to the app's evolution, such as enhancements to specific features or fixes to existing defects. This information can drive the development of the app and further improve the app. Apart from categorizing user reviews, researches \cite{r6, r14} have clustered similar user requirements and conducted a study on their priorities concerning various user demands. These researches aim to assist developers in determining which requirements should take precedence when developing the next version of the app.

Furthermore, in addition to the feedback information supplied by the general user base, the inherent information of the app itself constitutes a crucial source of data for researching the evolution and development of the app. Johann et al. \cite{r15} extracted features from app description information and user reviews, and then compared the extracted features to determine if each feature was mentioned in user reviews. Jiang et al. \cite{r16} constructed a classifier to extract features from app descriptions and then recommend missing features for similar apps. And Wu et al. \cite{r17} combined app description information and user reviews to identify key features highly correlated with app ratings, providing a means for developers to enhance the maintenance of their apps.

While the organized planning of the release schedule is the key to the success of an app. However, the rate at which the requirements of the general user are adopted by the app developers, referred to as ``user contribution" in this paper, may also provide a new perspective for developers to consider the evolution of the app, and provide guidance for the maintenance of the app in a healthy and orderly way. For instance, when user contribution is low in app evolution, whether developers should accept and adopt users' requirements and suggestions more when planning apps. And whether user stickiness will be provided when users' requirements can be responded to by developers in a fiercely competitive app store. To the best of our knowledge, no relevant work exists to quantitatively analyze the contribution of users to the evolution of an app. Therefore, the work in this paper aims to investigate the question of quantitative analysis of the contribution of general user requirements in the evolution of app development.

In this paper, we introduce a time-aware analysis approach based on user reviews and app update logs. This approach aims to extract interaction patterns between app developers and user feedback throughout the app evolution process, thus revealing the role of user participation in the evolution process of app development, which enables us to quantitatively mine the degree of user contribution in the app evolution. Specifically, first, based on the huge user review data, a deep learning algorithm is used to filter the user reviews, and reviews unrelated to the app features are disregarded, and further research is carried out only on the user reviews related to the app features. Then, user reviews related to app features are textually matched with the app update log to get reviews that are consistent with the new features or bug fixes in the update log, thus exploring the interaction between user feedback and app features. Finally, considering the temporal attributes inherent in user reviews and app update logs, a time-based feedback pattern of app features is derived to quantitatively analyze the contribution of user requirements in the evolution of the app.

To train and evaluate the approach proposed in this paper, we crawled and created a dataset based on the Apple App Store, including 2,178 update logs of 5 apps (4 Chinese apps and 1 English app) and 4,236,417 user reviews. The experimental results show that the approach proposed in this paper can effectively recognize irrelevant information in user reviews, as shown by the fact that the classifier is able to achieve an F1-Score of 92\%. In addition, this approach can effectively match the update logs and related user reviews with an average Precision of 90\% and an average Recall of 92\% for the five apps. Finally, relying on the matching results of update logs and user reviews, along with the temporal attributes of both, four feature feedback patterns are obtained, according to which it is found that the five apps studied in this paper have a maximum of 43.2\% of new features coming from the general user requirements during their developmental evolution, and a minimum of only 16.6\%.

In conclusion, the contribution of this paper is as follows:
\begin{itemize}
\item We propose a new analysis approach that integrates information from both users and apps to identify the interaction patterns between app features and user feedback throughout the development and evolution of apps, from which we mine user contribution, providing a new perspective for researching the evolution of apps.
\item We created a dataset from the Apple App Store consisting of 2,178 update logs for 5 apps and 4,236,417 user reviews.
\item We evaluated the performance of the proposed approach on the dataset, and the experimental results proved the effectiveness and practicality of the proposed approach in this paper.
\end{itemize}

The rest of the paper is organized as follows: section \ref{sec:II} describes the detailed process of this approach. Section \ref{sec:III} provides an overview of the dataset, along with the experimental design and analysis used to evaluate the approach. Section \ref{sec:IV} presents related work. Section \ref{sec:V} discusses potential threats to the findings in certain scenarios and outlines prospects for future research. Section \ref{sec:VI} concludes the paper.

\section{Approach}
\label{sec:II}
Apps in the app store contain information from two sources, one is the introductory descriptive information about the app from developers and vendors along with update logs informing users about new features in new versions, and the other is the information about users' reviews such as ratings and suggestions about the app, both of which continue to be made with the evolution of the app. In this paper, we consider valid user reviews as user requirements. The developer will take user feedback into account in the development plan during the evolutionary update of the app, and we consider the new features adopted in the update log of the new version that are required by users as user contributions. Therefore, it is feasible to quantitatively analyze the degree of user contribution in the evolution of app development. Taking this as a starting point, establishing a correlation between user requirements and developer responses, based on user reviews in the app store and the app's update logs, proves to be a promising approach for analyzing user contributions.

\begin{figure*}[htbp]
\centerline{\includegraphics[scale=0.46]{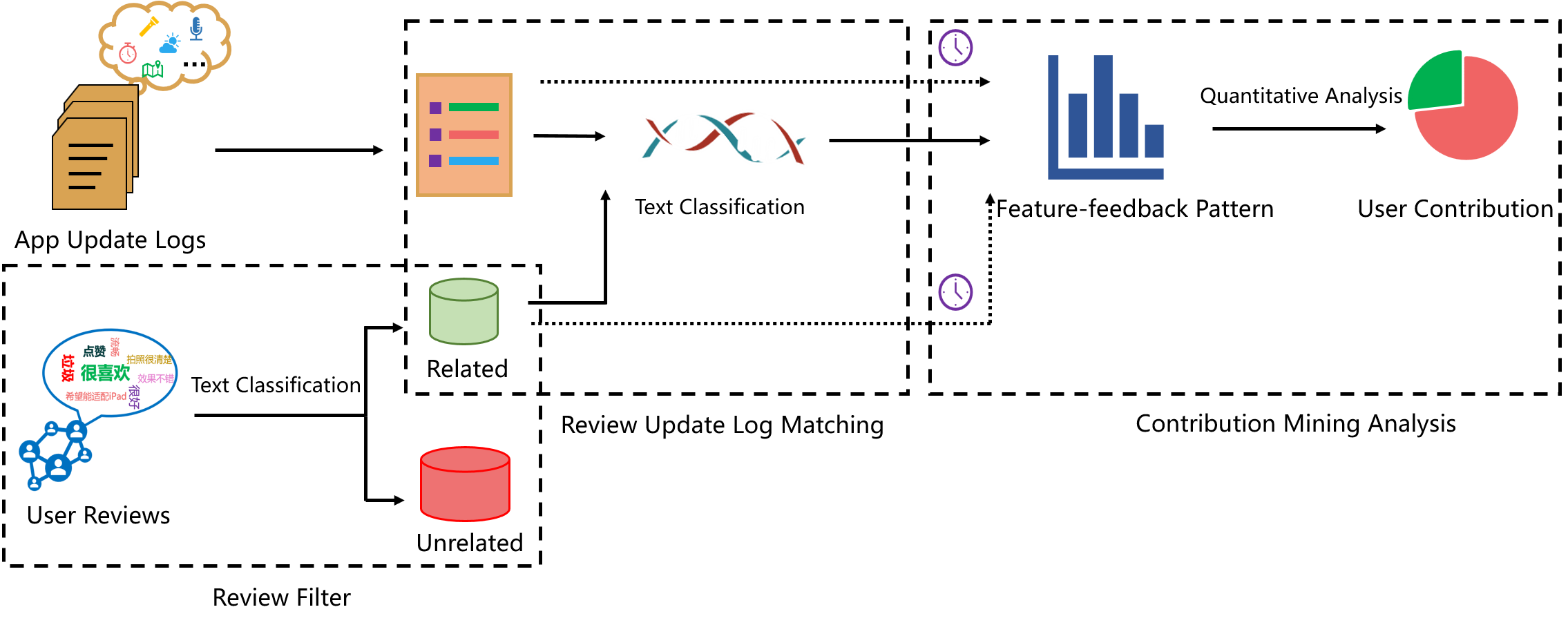}}
\caption{The overall of framework}
\label{fig1}
\end{figure*}

As previously mentioned, some popular apps may receive thousands of user reviews per day \cite{r5, r6}, and with app evolution continuing, review data may accumulate into the millions with varying quality of reviews, with only about one-third of user reviews being useful to developers \cite{r5, r6, r9}. Matching all the user reviews with the app update logs individually obviously is time consuming. To address this issue, this paper considers extracting valid reviews from massive user reviews firstly. To this end, this paper considers extracting valid reviews from massive user reviews firstly. Then, to obtain the update logs that have responded to user reviews, this paper e establishes a textual matching relationship between valid user reviews and app update logs through textual computation. Finally, to quantify user contribution from the intricate text data, this paper introduces the time dimension. Specifically, based on the time sequence of user reviews and update logs, the paper establishes the time-aware association between the user reviews and the update logs, which provides a scheme for quantifying the user contribution. The overall framework of this approach is shown in Figure~\ref{fig1} , comprising three main steps:

\textbf{(1) User Review Filtering:}  text classification algorithms are applied to filter information in user reviews that is not related to the app's features.

\textbf{(2) Review Update Log Matching:} For each update log of the app, textual representation modeling is performed to identify user reviews which match with it among the valid user reviews.

\textbf{(3) Contribution Mining Analysis:} Based on the matching of update logs and user reviews, taking the update time of each update log as the demarcation point, establish the time-aware association between user reviews and update logs, analyze the interaction pattern between app features and user feedback, and quantify the user contribution.

The specific steps of the approach are explained in detail below.

\subsection{Review Filter}
\label{sec:2.1}
\begin{figure}[htbp]
\centerline{\includegraphics[scale=0.4]{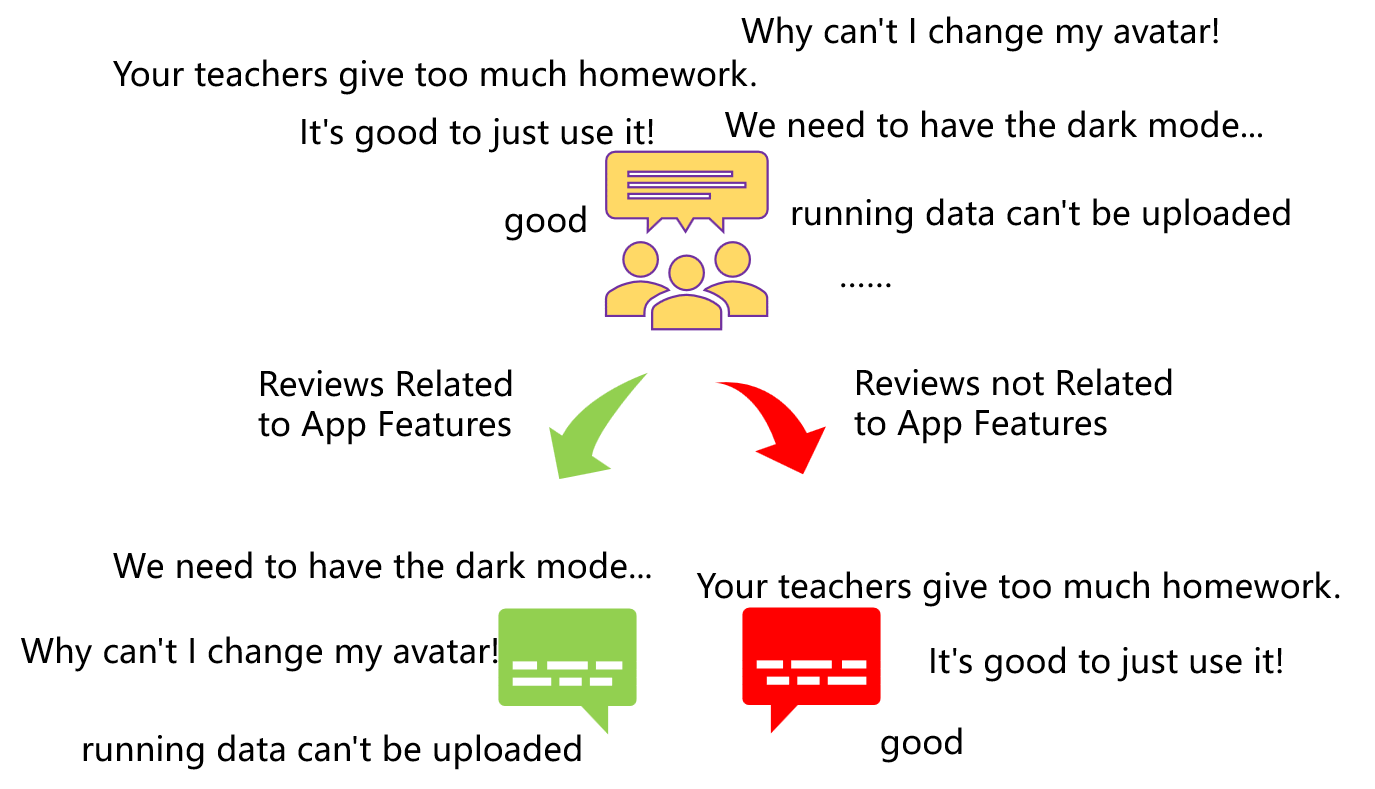}}
\caption{User Review Filtering}
\label{fig2}
\end{figure}

For many popular apps, the unstructured and varying quality of their user reviews makes identifying valuable information in those reviews a challenging task. Existing researches have done significant work on app review classification \cite{r5, r8, r10, r11, r12, r13}. However, the primary work of this paper is not to classify user reviews into fine-grained categories but rather to concentrate on extracting reviews that describe the features of the app or report bugs from the extensive dataset of user reviews. Specifically, reviews that contain descriptions of the app's features (including requests for new features, opinions of existing features), and errors that occur while using the app are considered relevant to the evolution of the app, as they inform developers of the features in the app that users are using and defects that may exist in the app. In contrast, reviews that express opinions about the app in general terms are considered irrelevant, as they only discuss overall user satisfaction without specifying a particular feature. Such reviews cannot be analyzed to ascertain the specific reasons behind the app attracting or repelling users. Therefore, in this paper, we classify user reviews into those related to app features (e.g., ``Why can't I change my avatar!", ``running data can't be uploaded", ``We need to have the dark mode...") and reviews unrelated to the app's features (e.g., ``good", ``It's good to just use it!" , ``Your teachers give too much homework."), i.e., filtering irrelevant reviews and extracting user reviews that are only evaluating, suggesting, or criticizing the app's features, as shown in Figure~\ref{fig2}. To preliminarily filter user reviews, this paper first preprocesses the review text, then trains a classifier using the real set, and finally employs the trained classifier to filter user reviews.

\subsubsection{Preprocessing}

In this paper, the pre-processing of user reviews for filtering involves eliminating noise from the reviews. User reviews include several parts such as title, time, rating, and content of the review, while user feedback is mainly focused on the content of the review. During the review filtering process, only the content of the review is processed in this paper. Note that, developers may also respond directly to user reviews\footnote{https://developer.apple.com/cn/app-store/product-page/}, expressing gratitude for downloads or guiding users to new features. According to the App Store's review mechanism, the above content will be appended to the corresponding user review content. However, the content of the developer's response may bring new information to the original user review, potentially causing a review that was initially unrelated to the app's features to now contain information relevant to the app's features. Therefore, this paper preprocesses user reviews by utilizing regular expressions to remove the content of developer responses within user reviews.

In addition, certain user reviews contain special symbols (e.g. \#, *, \^{}\_\^{}, \$-\$, etc.) and emoji hashtag symbols. To address this, regular expressions and the emoji toolkit\footnote{https://github.com/carpedm20/emoji} are employed to remove them. Note that, the special symbols removed in this paper don't contain ``@" because it has been observed that some features of the app use ``@" to indicate the mentioning of someone in a specific context. Finally, empty reviews that don't contain any characters after the preprocessing described above are removed in this paper.

\subsubsection{Classifier Training}

After user reviews are preprocessed, the review text is inputted into a deep learning classifier to train review classification models. BERT (Bidirectional Encoder Representations from Transformers) \cite{r18} has been pre-trained on a large number of datasets to learn generic syntactic and semantic features, and only needs to be fine-tuned on the specific dataset of the downstream task using the pre-trained model to make the model more adapted to the downstream task. And BERT has been proven to achieve significant results in a wide range of natural language processing tasks. Therefore, for the task of user review classification, this paper fine-tunes BERT on downstream tasks to train the classifier.

\begin{figure}[htbp]
\centerline{\includegraphics[scale=0.36]{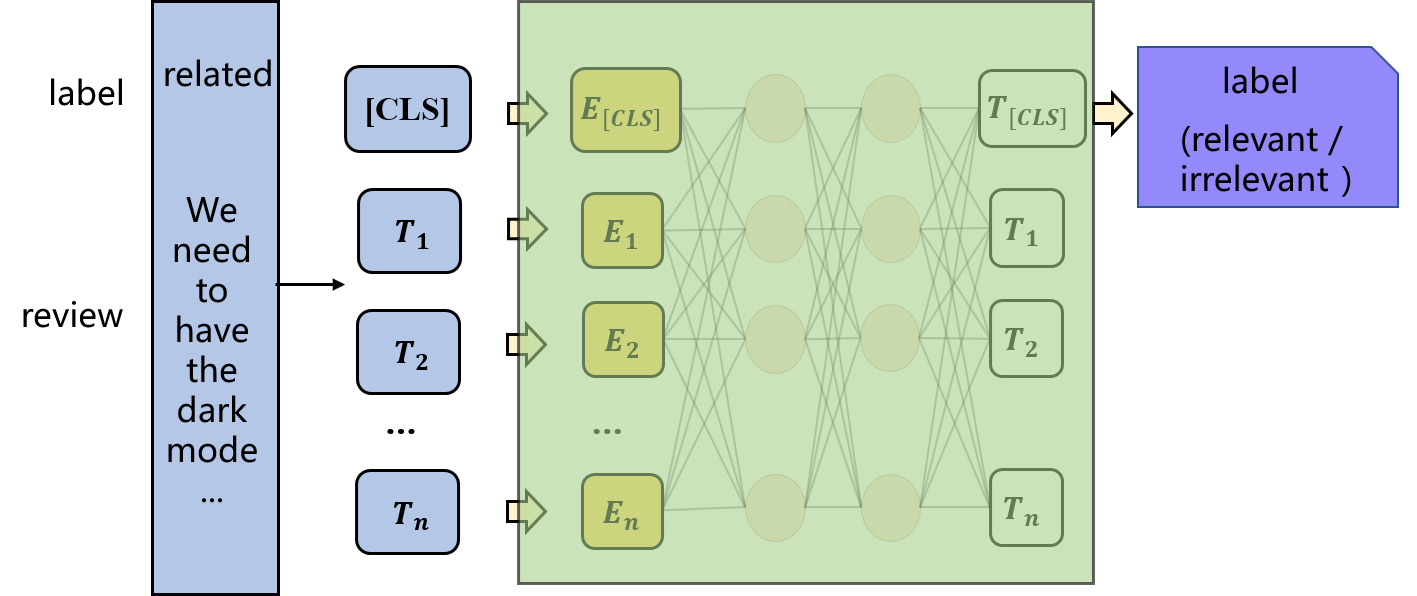}}
\caption{BERT-based Review Classification}
\label{fig3}
\end{figure}

Specifically, this is a binary classification task, where each data instance consists of a preprocessed review text and a manually labeled tag indicating whether the review is relevant to the app's feature (introduced in 3.2), inputted into the BERT model for fine-tuning, as shown in Figure~\ref{fig3}. For review texts in different languages, this paper adopts the pre-trained models (BERT-Base, Chinese and BERT-Base, Uncased)\footnote{https://github.com/google-research/bert} applicable to the corresponding languages for fine-tuning respectively.

Once the classifier is trained, all unlabeled reviews are predicted to be labeled by the classifier. Reviews identified as irrelevant to the app's features are disregarded in this paper, while reviews predicted to be relevant to the app's features are subsequently used to investigate whether they are related to the app's update logs.

\begin{figure}[htbp]
\centerline{\includegraphics[scale=0.6]{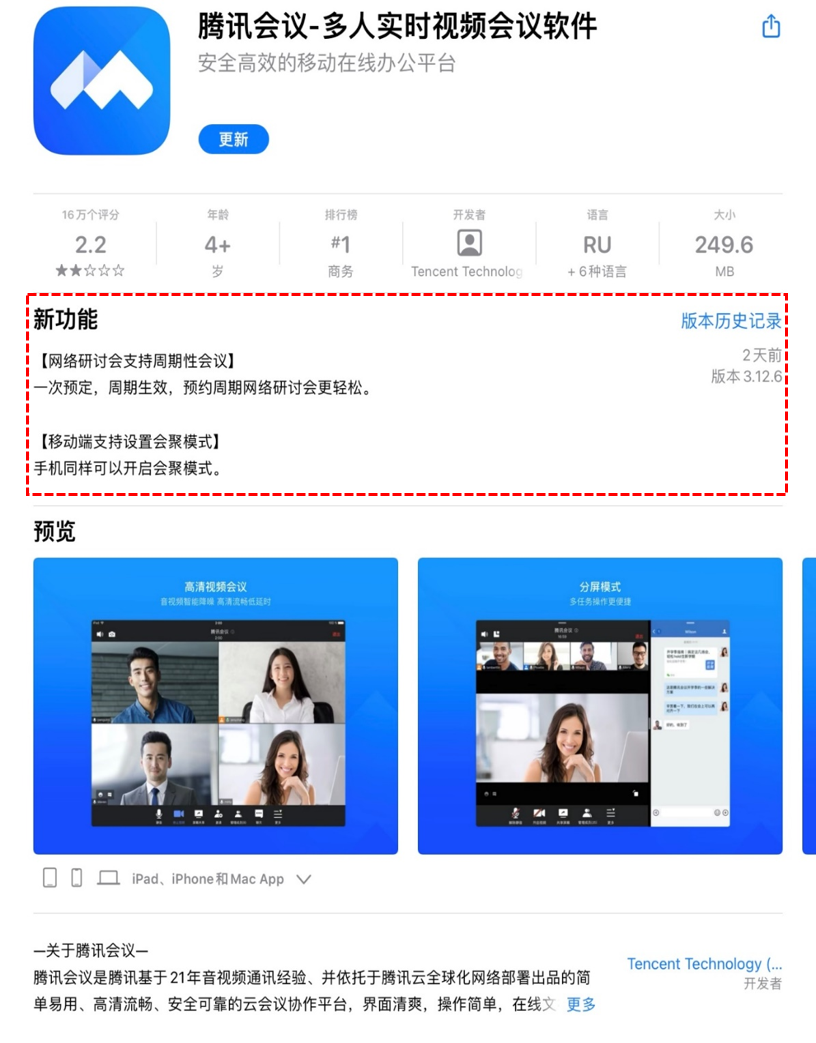}}
\caption{App Information in the App Store}
\label{fig4}
\end{figure}

\subsection{Review Update Log Matching}
\label{sec:2.2}
When the app development team releases a new version of the app, it records and informs users of what has changed in the current version through logs. In other words, after the app development team's internal planning or after the developers listen to users' opinions, they add a new feature or fix a certain bug to the app, and inform the users of the changes in general when releasing the new version, as shown in Figure~\ref{fig4}. This paper refers to the logs that developers release when introducing new versions and containing information about added features or bug fixes as ``update logs".

\subsubsection{Update Log Preprocessing}

\begin{figure*}[htbp]
\centerline{\includegraphics[scale=0.2]{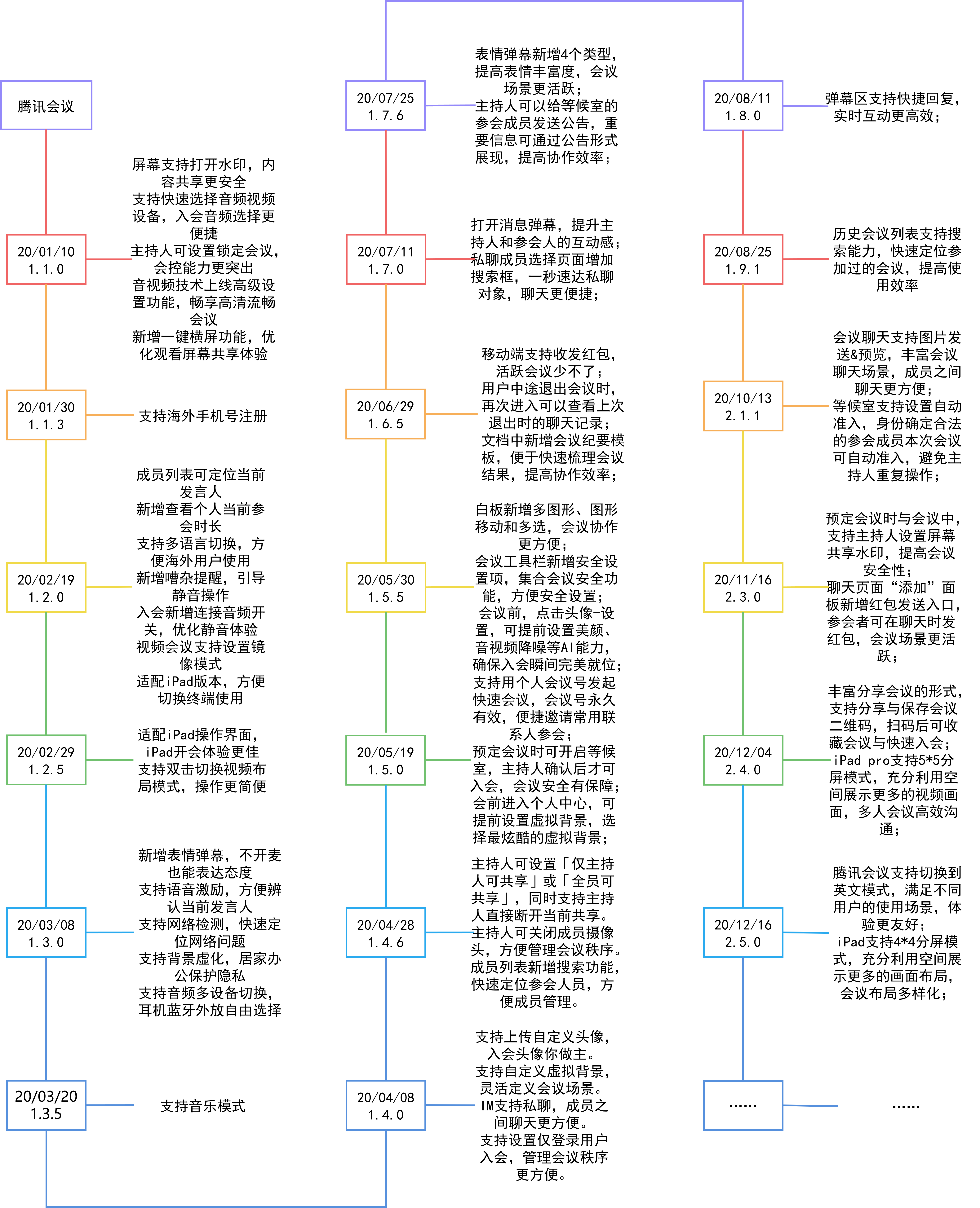}}
\caption{Tencent Meeting App update Log Timeline}
\label{fig5}
\end{figure*}

The release of a new version of an app typically involves updates to multiple aspects, resulting in multiple update logs. Nevertheless, some app update logs may not provide detailed information about the updates. For instance, an update log stating ``Fix some known bugs" lacks specific details about the bugs addressed, whereas an update log like ``Fix the bug of crash when clicking on the training plan" clearly informs the user that this update resolves the issue of the app crashing when clicking on the training plan. In addition, there are instances where the same app has two or more different versions with entirely identical update logs. In other words, the new version's update log followed the previous version's update log, and the current update does not inform users about the new features introduced. For the above two scenarios, we performed a preprocessing operation on the app update logs. All versions of the app were listed in the form of a timeline, including the update time, version number, and update logs. Subsequently, based on the timeline of the update logs: 1) Logs that did not describe the update information in detail were removed. Only complete descriptions explicitly detailing a specific modification item were retained, as update logs lacking detailed information could not provide valuable information; 2) In cases where multiple versions of an app shared the same update log, only the log with the lowest version was retained. because the developer has provided the new feature at an earlier time. Finally, app update log as shown in Figure~\ref{fig5} (using Tencent Meeting app as example).

\subsubsection{Matching User Reviews}

\begin{figure}[htbp]
\centerline{\includegraphics[scale=0.36]{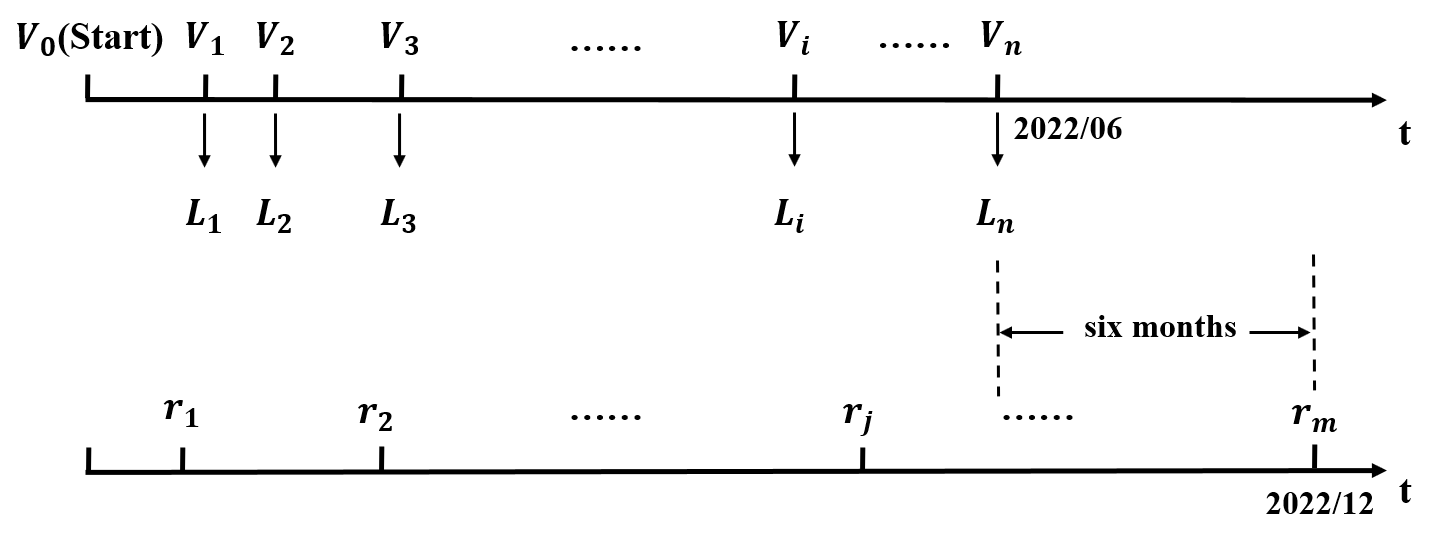}}
\caption{Review Log Matching}
\label{fig6}
\end{figure}

After the app update logs are preprocessed, the next step is to retrieve the user reviews describing a specific one of the update logs from the user reviews related to the app's features obtained after review filtering. Specifically, for an app, the evolution process and the process of user usage feedback is shown in Figure\ref{fig6}. Developers release the initial version $V_0$ of the app to the app store. At this stage, the app already possesses basic features, typically without accompanying update logs. Once the app is uploaded to the app store, users start to download and use the app. Consequently, they generate information such as user reviews describing suggestions for the app based on the experience of using the app, including the time of writing the review, which may contain defects found during use and requests for new features of the app. On a certain time, the developer releases version $V_1$ and uploads the corresponding update log $L_1$ to inform users of the changes made in this version. Users update to version $V_1$ and continue to use it, concurrently offering feedback to the app developer through reviews of their experience and suggestions.

Each release of a new version $V_i$ of the app is accompanied by a corresponding update content $L_i$, and each update content $L_i$ consists of multiple update logs $l_{ip}$, denoted as $L = \{L_1, L_2, ..., L_n\}$, $L_i = \{L_{i1}, L_{i2}, ..., L_{ip}\}$ where $1 \le i \le n$, $n$ is the number of updates, and $p$ is the number of update logs for a particular update. For user reviews, $R = \{r_1, r_2, ..., r_m\}$, and $m$ is the total number of user reviews for the app. For each update log $l_{ip}$ of the app, retrieve the user reviews in the set of user reviews that are related to it, i.e., the user review describes the update log. Obtain the set of matches between update logs and user reviews, denoted as $\{L_{ip}, r_j \in (L, R)\}$. Special notice is that the user reviews considered in this paper are from a period six months longer than the update logs. This is because after the release of the latest version of the app, users may not immediately update to the latest version. Therefore, there is a certain lag in user perception, and there is insufficient data for reviews. A period of time needs to be set aside to receive feedback from the users. In addition, this paper does not employ text patterns to extract feature phrases from the update logs as in \cite{r15, r16, r17}, and then match them with user reviews. e.g., extracting ``open watermark" from ``open watermark on screen support for safer content sharing", and then identifying that mention ``open watermark". The reason for not using text patterns to extract feature phrases is that, as the app evolves iteratively, developers optimize the app for the same feature across different user roles and usage scenarios. Extracting feature phrases through text patterns may not capture the full update logs information, which is crucial for informing users about new features in the latest version. For instance, the text patterns in \cite{r17} can extract feature phrases such as ``Remove Members", ``Allow Entry", etc., from the update logs ``When removing a member, you can set the setting to not allow the member to rejoin the meeting, which makes the meeting control more convenient". However, the approach of using text patterns to extract feature phrases may lead to a lack of generalization for new features in the update log. Additionally, multiple distinct update logs could be associated with the same feature phrases, resulting in the loss of temporal properties of the update log. Therefore, this paper utilizes the entire update log for matching, ensuring a more comprehensive and accurate representation of the evolution of app features over time.

In the process described above, determining whether a user review describes the update log can be framed as a semantic matching relationship between the update log $l_{ip}$ and arbitrary user review $r_j$. To achieve semantic matching between update logs and user reviews, this paper fine-tunes the BERT pre-trained model for the downstream task of sentence pair classification. This task involves providing an update log $l$, a user review $r$, and a binary label $f$ indicating whether the user review $r$ describes a new feature provided by the update log $l$ (1 indicates a match, 0 indicates not). The trained BERT model then evaluates the semantic similarity between any given update log $l_{ip}$, and user review $r_j$. The evaluation is based on whether a given set of update logs matches with a user review, represented as:
$$M_{l_{ip}r_j} = \begin{cases}1 &\text{ } p \ge 0.5 \\ c &\text{ } p \le 0.5\end{cases}$$ 
where 1 indicates that the update log $l_{ip}$ and the user review $r_j$ are semantically similar. In other words, the update log responds to the user review, or the user review describes the content of the current update log, and 0 indicates the opposite. The final set $\{(L_{ip}, R_j) \in (L, R)\}$ is derived for each update log matching a user review.
\subsection{Contribution Mining Analysis}

\begin{figure}[htbp]
\centerline{\includegraphics[scale=0.4]{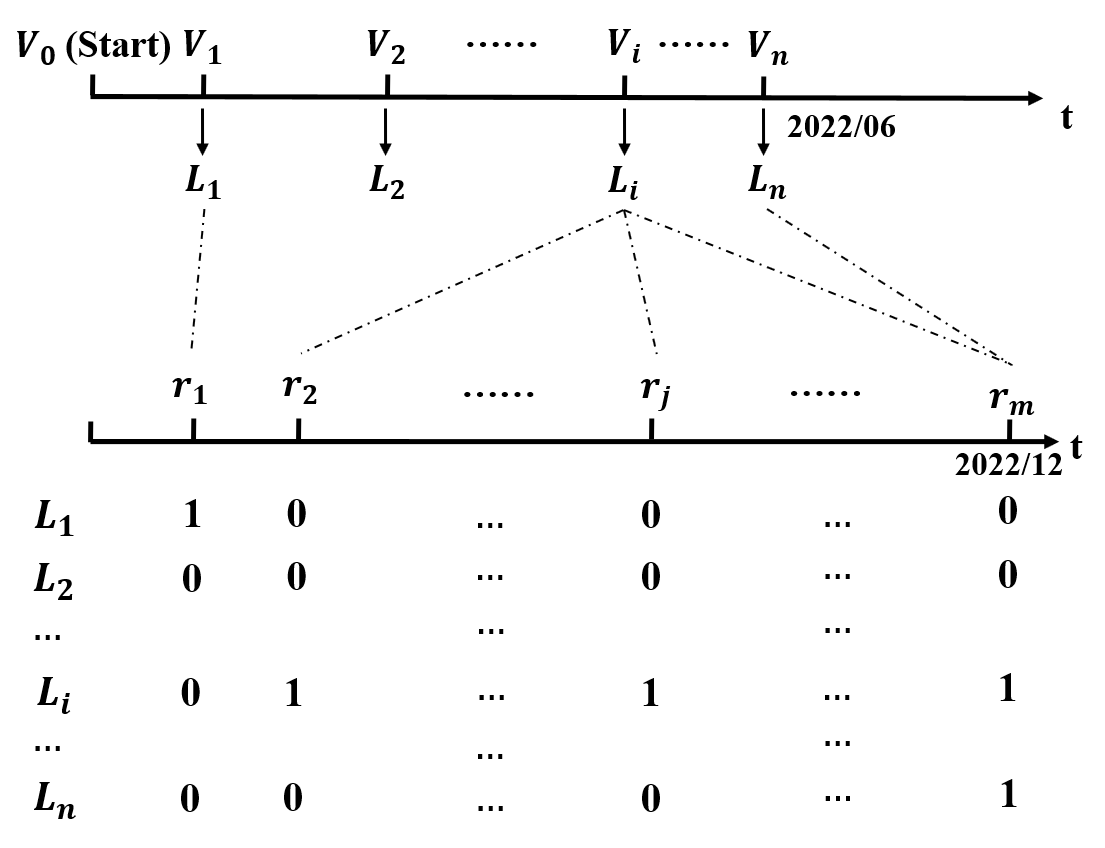}}
\caption{The overall of framework}
\label{fig7}
\end{figure}

\subsubsection{Feature Feedback Pattern}
In the two preceding sections, following the procedures of review classification and semantic matching, matching relationships were derived for all sentence pairs within the context of ``Update Logs -  User Reviews". Obviously, if multiple users submit identical feature suggestions or error reports within the same timeframe, a single update log entry may correspond to multiple user reviews. Building upon the efforts outlined in the previous two sections, for any update $L_i$ where $i = (1, ..., |L_n|)$, within the evolutionary cycle of the app, the matching status with associated user reviews is depicted in Figure\ref{fig7}.

Based on the matching results, we can establish a correspondence between the app update logs and user reviews, enabling us to infer the interaction between app feature and user feedback. In this paper, we define this interaction as a ``feature feedback pattern" and aims to extract user contributions from these feature feedback patterns.

In this paper, feature feedback patterns based on the app update time. Firstly, according to the matching results between update logs and user reviews, for any update log $l_{ip}$, when $M_{l_{ip}r_j} = 1$, it indicates that user review $r_j$ is associated with it. Secondly, based on the overall matching relationship between update logs and user reviews, for each update log, we use its update time as a starting point to categorize the associated user reviews. Specifically, user reviews submitted prior to the update log's update time are defined as ``pre-update reviews", whereas those submitted subsequent to the update log's update time are considered ``post-update reviews". Finally, by utilizing the update time of the update log as a baseline and considering the quantities of pre-update reviews and post-update reviews, we establish an interaction relationship between feature and feedback based on time. This relationship is denoted as a ``feature feedback pattern".

\subsubsection{User Contribution Mining}

In the previous section, the definition of feature feedback patterns during the evolution of the app was explained. This section will outline how we extract user contributions to the evolution of the app through these feature feedback patterns.
Based on the definition of feature feedback patterns as described earlier, four types of patterns can be identified: 1) Update logs with no associated user reviews both before and after the update (0-0 pattern). 2) Update logs with no associated user reviews before the update but with related user reviews after the update (0-1 pattern). 3) Update logs with associated user reviews before the update but without related user reviews after the update (1-0 pattern). 4) Update logs with associated user reviews both before and after the update (1-1 pattern). We analyze the reasons behind the formation of different patterns.

Update logs with no associated user reviews both before and after the update (0-0 pattern): The reason for this pattern is that users did not have a requirement for the specific feature when using the app. Even if the app introduced this feature, users did not experience any noticeable change, or the feature didn't capture users' attention. In other words, there were no user reviews expressing their opinions on this feature throughout its development. We suggest that such features are part of the internal planning by developers, leading to the formation of this type of feature feedback pattern.

Update logs with no associated user reviews before the update but with related user reviews after the update (0-1 pattern): A parallel situation to the previous 0-0 pattern is observed, wherein users initially did not express a demand for this specific feature. However, in contrast to the 0-0 pattern, the introduction of the new feature generated user interest. Following the feature's release, it captured users' attention, prompting them to share their experiences with the new feature and offer suggestions for its improvement. This pattern is also attributed to planning by developers.

Update logs with associated user reviews before the update but without related user reviews after the update (1-0 pattern): In contrast to the previous two patterns, this particular pattern signifies the presence of user reviews associated with the update log before the update. Initially, the app did not possess the current feature, but developers responded to user feedback by incorporating a new feature. Following the app update, the user requirements were satisfied, which explains the absence of new reviews pertaining to this feature. This pattern is attributed to user-driven requirements for specific features.

Update logs with associated user reviews both before and after the update (1-1 pattern): This pattern is distinguished from the 1-0 pattern in that even after the feature update, users continue to provide feedback on the feature. Likewise, these features are driven by users' initial requirements, which is why this pattern forms.

Based on the analysis of the reasons for the formation of these four types of patterns, it is suggested that the features within the (0-0) and (0-1) patterns result from developer planning. Initially, users did not express a requirement for these features, and developers provided them before user requirement emerged. Therefore, these patterns are categorized as developer-driven patterns. For patterns (1-0) and (1-1), they are based on user requirements, where user feedback precedes the update and the developers subsequently implement the updates, it indicates that user demands were expressed first, and developers subsequently implemented the updates. Therefore, these patterns are categorized as user-driven patterns.

Finally, through the implementation of statistical analysis across all feature feedback patterns, we calculate the proportion of new features that were initiated by user requirements in the app's development since its release. This proportion is used to determine the extent of user contributions in the evolution of the app.

\section{Experimental Setup and Analysis}
\label{sec:III}
In this section, we initially outline the research questions we aim to investigate and then introduce the detailed methods employed for evaluation.

\subsection{Research Questions}
In this paper, we put forth the following three research questions.

\textbf{RQ1} \textit{To what extent is the approach for filtering user reviews effective? This research question primarily evaluates the accuracy of our approach in classifying user reviews, as described in section \ref{sec:2.1}.}   %  as described in the methodology section detailing the user review filtering process.

\textbf{RQ2} \textit{How effective is the approach in matching app update logs with their associated user reviews? This research question primarily focuses on the accuracy of our approach in automatically identifying user reviews related to the content of each update log, as described in  section \ref{sec:2.2}.}   % as described in the methodology section detailing the user review matching process.

\textbf{RQ3} \textit{What is the varying degree of user contributions in different apps? This research question primarily explores the extent to which new features in the app's developmental process originate from user suggestions.}

\subsection{Dataset}
The primary app stores for mainstream mobile platforms include the iOS-based App Store and the Android-based Google Play. The App Store serves as a unified platform exclusively for app downloads on iOS devices, features a standardized system for reviews and ratings. The dataset employed in this study is obtained from the App Store in both mainland China and the United States.

To create the dataset, we initially conducted a preliminary screening of apps within various categories on the Apple App Store in mainland China. The top three apps from each category's rankings were then chosen as candidate apps. To ensure an adequate amount of data, encompassing both user reviews and update logs, for training classifiers, we further refined the candidate apps based on the following criteria: 1. The app no fewer than 10,000 user reviews. 2.The app no fewer than 100 update logs. Finally, we randomly selected four Chinese apps (Tencent Meeting, TikTok, DingTalk, and Keep) and an additional English app (Zoom) from the US App Store. The data for the five apps comprises 2,489 update logs and 4,237,031 original user reviews, utilized to address the research questions.

\subsection{Evaluation Methodology}
This section will address the three research questions mentioned earlier.

\subsubsection{Methodology for Addressing RQ1}
\label{sec:3.3.1}

The first research question explores the effectiveness of the proposed approach in classifying original user reviews into those related and unrelated to app features. To address this, a real dataset for review classification needs to be created.

To create the real dataset, a process similar to that outlined in \cite{r3} was adopted. During the real dataset creation process, annotators systematically evaluated the content of user review samples based on annotation guidelines. For each review, three annotators independently assessed whether the content was related to app features, unrelated, or labeled as an additional category, ``uncertain", indicating that the annotators could not make an independent judgment, a allowing for subsequent discussion in the real dataset confirmation stage. The creation process of the real dataset involves four steps: 1) annotation guidelines design, 2) review sampling, 3) annotating user review samples, and 4) confirming the real dataset. Each step is described as follows.

1) Annotation Guidelines Design: We designed annotation guidelines that described our understanding of whether reviews were related to app features. This was done to systematize the task of creating the real dataset and minimize disagreements among annotators. The guidelines included instructions about the annotation task, clear definitions, and examples of the classification approach. Additionally, it provided brief descriptions of each app.

2) Review Sampling: User reviews often vary in their wording. Because of the diversity of individual users and the unique characteristics of user reviews, even for the same app at different time periods, there can be significant variations. To represent the review data comprehensively, we randomly sampled reviews from each app, resulting in 93,863 user reviews for annotation.

3) Review Annotation: Three graduate students with experience in app development were tasked with annotating the sampled reviews. Before commencing the annotation task, all annotators were required to acquaint themselves with the annotation guidelines and engage in detailed discussions about each instance during a meeting. Throughout the annotation process, annotators independently analyzed the content of reviews and labeled them as either related or unrelated to app features. When annotators were uncertain about the type of a particular review, it was marked as ``uncertain" for further determination in subsequent steps.

4) Real Dataset Confirmation: After all the reviews to be annotated were labeled by the three annotators, we collected all the reviews with inconsistent labels and those marked as ``uncertain". We invited two doctoral students to join a meeting discussion to resolve any ambiguities among the annotators. In the end, 9,186 reviews related to app features and 84,677 reviews unrelated to app features were confirmed among the annotated reviews, as shown in Table~\ref{tab1}.

\begin{table}[tbp]
\caption{Annotated Reviews}
\begin{center}
\scalebox{1.2}{
\begin{tabular}{cccc}
% sanxianbiao
\toprule    % dingbu cuxian
App             & Annotated & Relevant & Irrelevant \\
\midrule    % zhongjian xixian
Tencent Meeting & 5309      & 1699     & 3610       \\
TikTok          & 63834     & 1480     & 62354      \\
DingTalk        & 10846     & 3283     & 7563       \\
Keep            & 6829      & 1744     & 5085       \\
Zoom            & 7045      & 980      & 6065       \\
Total           & 93863     & 9186     & 84677     \\
\bottomrule % dibu cuxian
\end{tabular}
}
\label{tab1}
\end{center}
\end{table}

In this paper, 70\% of the user review data was randomly selected as the training dataset for fine-tuning the BERT classifier employed for review classification. Additionally, 20\% was utilized as a validation dataset for parameter tuning, while the remaining 10\% was reserved to evaluate the performance of the classifier. Furthermore, the paper employed the commonly used text classification algorithm Naive Bayes \cite{r8, r11} as a baseline method. Precision, Recall, and F1-score were employed to assess the classification performance of the model:
$$Precision = \frac{TP}{TP + FP}$$
$$Recall = \frac{TP}{TP + FN}$$
$$F1-Score = \frac{2 \times Precision \times Recall}{Precision + Recall}$$
where $TP$ represents true positives (correctly predicted positive class), $FP$ represents false positives (negative class predicted as positive), and $FN$ represents false negatives (positive class predicted as negative).

This research question assessed the effectiveness of the approach in the classification of user reviews. Table~\ref{tab2} reports the achieved Precision (P), Recall (R), and F1-score (F1) in filtering user reviews. These results indicate that the approach can effectively and accurately filter selected app reviews. On average, for user reviews related to app features, the approach achieves a precision of 0.91, recall of 0.91, and an F1-Score of 0.91. This marks an improvement of 0.10, 0.08, and 0.09, respectively, compared to the traditional baseline method (Naive Bayes). For user reviews unrelated to app features, it achieves an average precision of 0.92 (+0.05), recall of 0.91 (+0.07), and an F1-Score of 0.92 (+0.07).

Answer to RQ1: The approach effectively filters user reviews, achieving an average F1-Score of 0.91 for classifying relevant user reviews and an F1-Score of 0.92 for classifying irrelevant reviews.

\begin{table}[tbp]
\caption{The Result of Review Filtering}
\begin{center}
\begin{tabular}{cccccccc}
\toprule
\multirow{2}{*}{Method}     & \multirow{2}{*}{App} & \multicolumn{3}{c}{Relevant} & \multicolumn{3}{c}{Irrelevant} \\
                            &                      & P      & R          & F1     & P        & R        & F1       \\
\midrule
\multirow{6}{*}{NB}         & Tencent      & 0.73   & 0.81       & 0.77   & 0.81     & 0.77     & 0.79     \\
                            & TikTok               & 0.84   & 75630.84   & 0.84   & 0.91     & 0.82     & 0.86     \\
                            & DingTalk             & 0.80   & 0.84       & 0.82   & 0.90     & 0.83     & 0.86     \\
                            & Keep                 & 0.79   & 0.76       & 0.78   & 0.84     & 0.84     & 0.84     \\
                            & Zoom                 & 0.88   & 0.88       & 0.88   & 0.91     & 0.93     & 0.92     \\
                            & Avg                  & 0.81   & 0.83       & 0.82   & 0.87     & 0.84     & 0.85     \\
\midrule
\multirow{6}{*}{This } & Tencent      & 0.88   & 0.84       & 0.86   & 0.88     & 0.89     & 0.89     \\
                            & TikTok               & 0.93   & 0.92       & 0.92   & 0.96     & 0.92     & 0.94     \\
                            & DingTalk             & 0.91   & 0.91       & 0.91   & 0.90     & 0.93     & 0.91     \\
                            & Keep                 & 0.90   & 0.90       & 0.89   & 0.89     & 0.91     & 0.90     \\
                            & Zoom                 & 0.93   & 0.99       & 0.96   & 0.99     & 0.92     & 0.96     \\
                            & Avg                  & 0.91   & 0.91       & 0.91   & 0.92     & 0.91     & 0.92    \\
\bottomrule
\end{tabular}
\label{tab2}
\end{center}
\end{table}

\subsubsection{Methodology for Addressing RQ2}

The second research question explores the feasibility of accurately extracting user reviews that describe update logs. In this study, we extracted the update logs for the previously mentioned five apps. After preprocessing to remove update logs that did not clearly specify update information and removing duplicate entries, the number of update logs for the five apps is shown in Table~\ref{tab3}.

Creating the real dataset for matching user reviews with app update logs involves annotating the data to establish the relationships between user reviews and app update logs. In the dataset obtained for this study, the volume of user reviews significantly exceeds that of update logs. Therefore, it is more feasible to sample and annotate update logs for matching with sampled user reviews. Note that, this task is different from the previous step of review filtering. The review filtering annotation task entails analyzing whether user reviews describe app features, whereas this task involves determining whether user reviews are related to specific feature described in app update logs. With hundreds of app update logs and tens of thousands of user reviews, the combination of all update logs and user reviews reaches a massive scale (149 * 3549710 = 528906790, the number of update logs for the TikTok app multiplied by the number of user reviews). It is impractical to manually analyze and label the matching relationships between user reviews and update logs.

Therefore, to construct the real dataset, we initially consider the word similarity between user reviews and update logs. User reviews and update logs with high word similarity are initially designated as positive samples. In text similarity algorithms, cosine similarity measures the size of the difference between two texts by calculating the cosine value of the angle between two vectors in a vector space \cite{r19}. In this context, we employ the cosine similarity algorithm to calculate the word similarity between update logs and user reviews related to app feature. The update logs randomly selected from each app are individually computed with the 9,186 user reviews confirmed as related to app features in section \ref{sec:3.3.1}. Specifically, let $D_{s1}$ and $D_{s2}$ be the set of terms occurring in update log $l_{ip}$ and user review $r_j$. Define $T$ as the union of $D_{s1}$ and $D_{s2}$, and let $t_i$ be the $i^{th}$ element of $T$. Define the term frequency vectors of $l_{ip}$ and $r_j$ as:
$$l_{ip}^{TF} = [nl_{ip}(t_1), nl_{ip}(t_2), ..., nl_{ip}t_N]$$
$$r_j^{TF} = [nr_j(t_1), nr_j(t_2), ..., nr_jt_N]$$
where $nl_{ip}(t_i)$is the number of occurrences of term $t_i$ in $l_{ip}$. Therefore, the cosine similarity between the update log and the user reviews is expressed as follows:
$$Sim\_Cosine = \frac{l_{ip}^{TF} \cdot r_j^{TF}}{\begin{Vmatrix} l_{ip}^{TF} \end{Vmatrix} + \begin{Vmatrix} r_j^{TF} \end{Vmatrix}}$$
where the dot product $\cdot$ is the scalar product and norm $\begin{Vmatrix} \end{Vmatrix}$ is the Euclidean norm. We set a threshold of 0.3, meaning that when the $Sim\_Cosine$ between a user review and an update log is greater than or equal to 0.3, it is considered a match and categorized as a positive sample, indicating that the user review $r$ describes the update log $l$.

\begin{table}[tbp]
\caption{Number of App Update Logs}
\begin{center}
\begin{tabular}{ccccccc}
\toprule
App     & Tencent & TikTok & DingTalk & Keep & Zoom & Total \\
\midrule
Numbers & 132             & 149    & 827      & 463  & 607  & 2178 \\
\bottomrule
\end{tabular}
\label{tab3}
\end{center}
\end{table}

\begin{figure*}[htbp]
\centering
\subfigure[]
{
    \begin{minipage}{.45\linewidth}
        \centering
        \includegraphics[height = 6cm, width = 8cm]{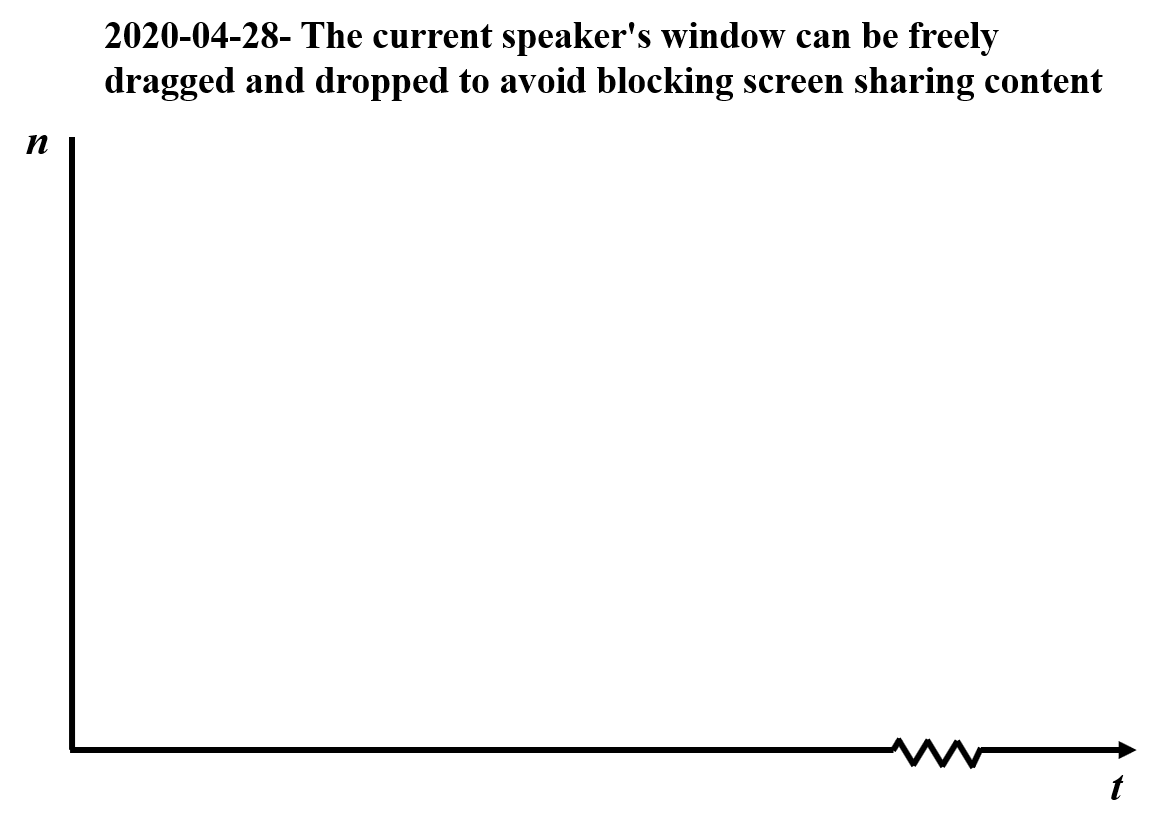}
    \end{minipage}
}
\subfigure[]
{
    \begin{minipage}{.45\linewidth}
        \centering
        \includegraphics[height = 6cm, width = 8cm]{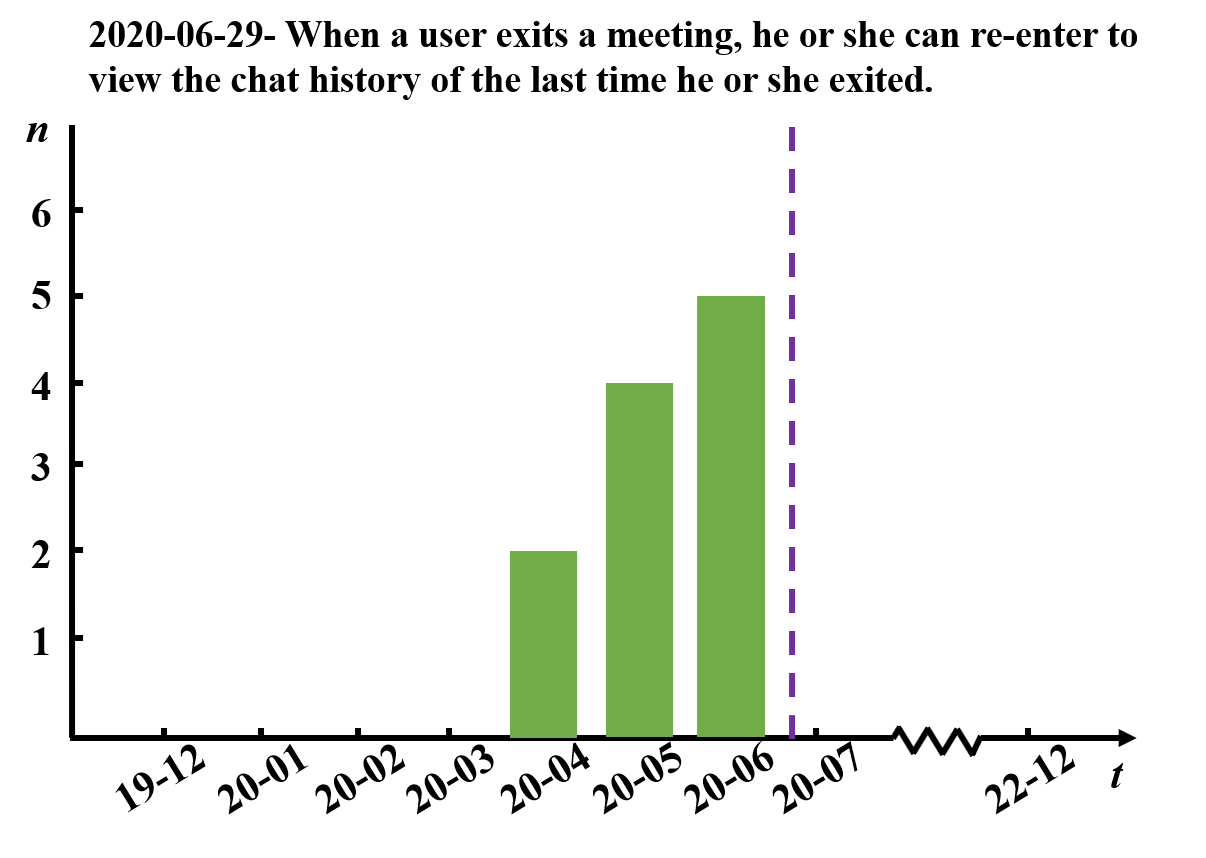}
    \end{minipage}
}
\subfigure[]
{
    \begin{minipage}{.45\linewidth}
        \centering
        \includegraphics[height = 6cm, width = 8cm]{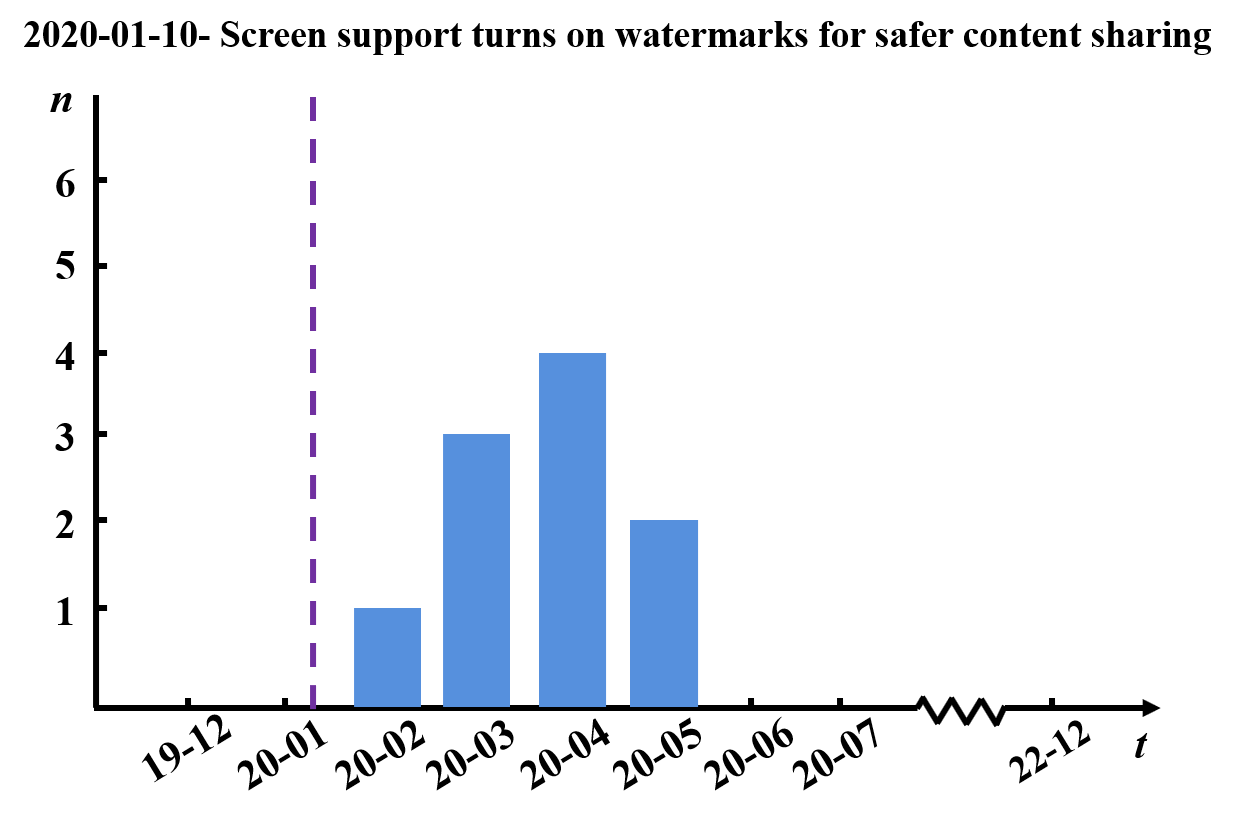}
    \end{minipage}
}
\subfigure[]
{
    \begin{minipage}{.45\linewidth}
        \centering
        \includegraphics[height = 6cm, width = 8cm]{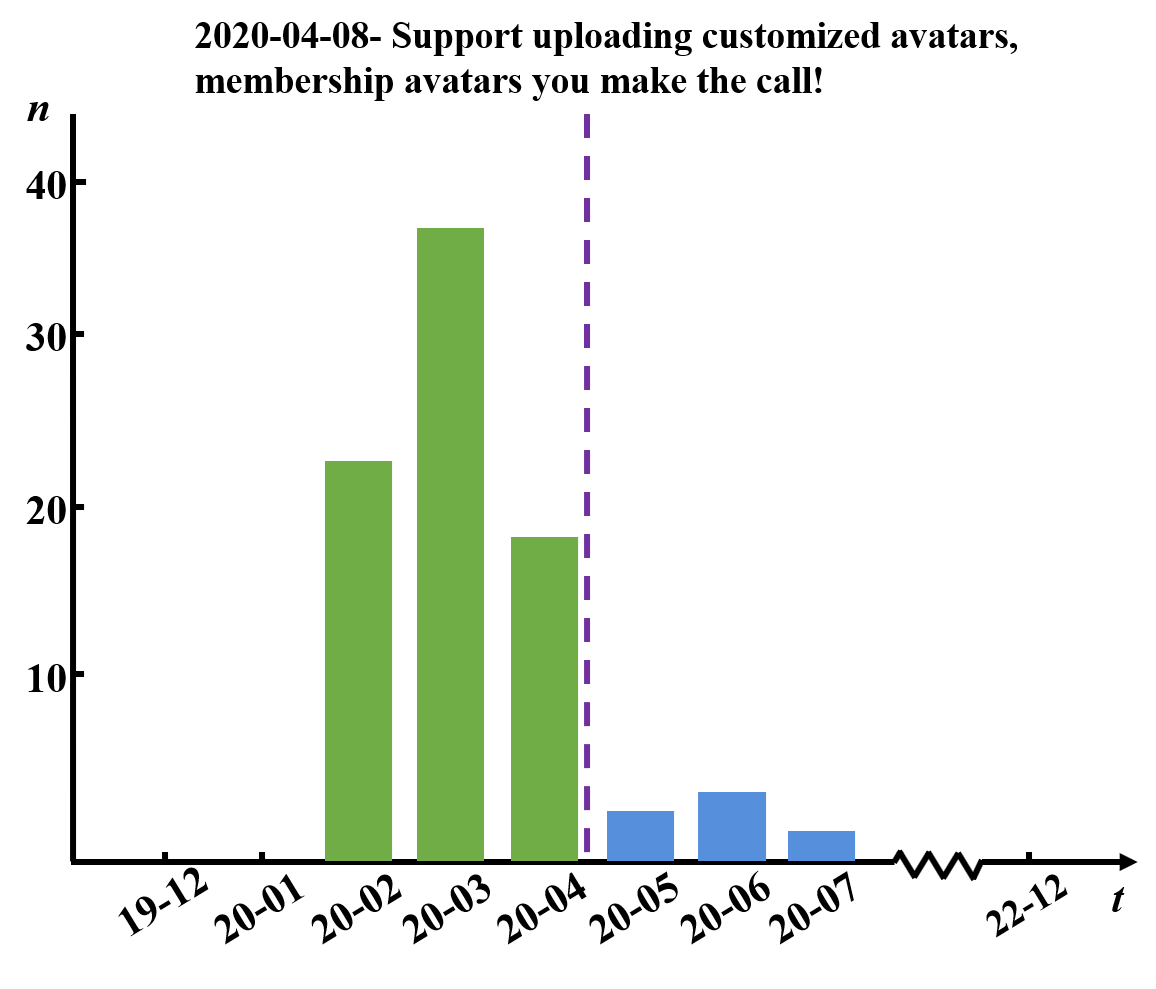}
    \end{minipage}
}
\caption{Four types of feature feedback patterns}
\label{fig8}
\end{figure*}

After the preliminary calculation using cosine similarity, the matching data only demonstrated consistency in word frequency similarity and did not genuinely reflect semantic similarity between sentence pairs. Therefore, manual verification was necessary for sentence pairs that were initially identified as similar. For example, the update log ``Perfectly compatible with iOS 10" and the user review ``Perfectly compatible with iPad, 10/10" had a cosine similarity of 0.58. It is evident that the user review does not describe the new feature in the current update log, and the two are not semantically similar. Consequently, this paper manually verified and corrected the data initially identified as positive samples. Guidelines for the task of correcting matching between update logs and user reviews were established. Three graduate students with over one month of experience using the labeled app performed the correction tasks following the established guidelines. After calculating word similarity using cosine similarity and manual verification, the final positive sample situations for matching update logs and user reviews for each app are shown in Table~\ref{tab4}.

Having obtained the positive samples (i.e., sentence pairs where update logs and user reviews are semantically similar), an equal number of negative samples (i.e. sentence pairs where update logs and user reviews are not semantically similar) were randomly sampled for fine-tuning the BERT sentence pair classification task. For each app, the positive and negative samples were randomly split into 70\% training data, 20\% validation data, and 10\% testing data. Precision, Recall, and F1-Score were calculated accordingly to evaluate the method's matching effectiveness.

\begin{table}[tbp]
\caption{Update Log and User Review Matching Dataset}
\begin{center}
\begin{tabular}{ccccccc}
\toprule
App             & Logs & Reviews & Positive Samples \\
\midrule
Tencent Meeting & 60          & 1699            & 636              \\
Tiktok          & 68          & 1480            & 532              \\
DingTak         & 411         & 3283            & 338              \\
Keep            & 152         & 1744            & 170              \\
Zoom            & 309         & 980             & 183              \\
Total           & 1000        & 9186            & 5979            \\
\bottomrule
\end{tabular}
\label{tab4}
\end{center}
\end{table}

\begin{table}[tbp]
\caption{The Results matching }
\begin{center}
\scalebox{1.2}{ % add here
\begin{tabular}{ccccccc}
\toprule
App             & P    & R    & F1   \\
\midrule
Tencent Meeting & 0.67 & 0.73 & 0.70 \\
Tiktok          & 0.99 & 0.98 & 0.99 \\
DingTalk        & 0.97 & 0.98 & 0.97 \\
Keep            & 0.98 & 0.97 & 0.97 \\
Keep            & 0.98 & 0.97 & 0.97 \\
Zoom            & 0.94 & 0.94 & 0.94 \\
Avg             & 0.90 & 0.92 & 0.91 \\
\bottomrule
\end{tabular}
}   % add \here
\label{tab5}
\end{center}
\end{table}

Table~\ref{tab5} presents the achievable Precision, Recall, and F1-Score. It can be observed that the approach can accurately extract user reviews related to the features of each app. Specifically, the performance of user reviews matching update logs on the test set for the five apps is as follows: Precision is at least 67\%, with an average of 90\%. In addition, the average Recall reaches 92\%, high F1-Scores are achieved as well.

Answer to RQ2: This approach is highly precise in matching app update logs with their related user reviews, with an average F1-Score of 91\% for user review matching.

\subsubsection{Methodology for Addressing RQ3}

The primary objective of this paper is to investigate the contributions of users to diverse app developments. Following the validation of steps one and two through experiments all user reviews were filtered and matched with the corresponding app update logs, resulting in the matching situations for each app's every update log with user reviews. After pattern analysis, some of the feature feedback patterns are shown in Figure~\ref{fig8}. The upper part displays the time of the update log and a specific update log. The horizontal axis represents the time period, the vertical axis represents the number of user reviews related to the update log, and the purple dashed line on the horizontal axis represents the time of the update log. Figure~\ref{fig8}(a) represents the 0-0 pattern, denoting there are no user reviews related to the app's evolution. Figure~\ref{fig8}(b) represents the 1-0 pattern of update logs, indicating that there are user reviews related to the update log before it, but none after the app update. Figure~\ref{fig8}(c) represents the 0-1 pattern of update logs, signifying that user reviews related to the update log only emerge after its release. Figure~\ref{fig8}(d) represents the 1-1 pattern of update logs, indicating the existence of user reviews related to the update log both before and after the update.

The quantification of the four types of feature feedback patterns for each app is shown in Table~\ref{tab6}. It was found that for all the update logs analyzed in this paper for the five apps, 56\%-83\% of the new features were planned internally by the developers, while approximately 16\%-43\% of the new features were driven by user requirements. To some extent, users have influenced the evolution of the apps, but the majority of features were planned by the developers.

\begin{table}[tbp]
\caption{Summary of Feature Feedback Patterns}
\begin{center}
\begin{tabular}{ccccccc}
\toprule
App             & 0-0 & 0-1 & \begin{tabular}[c]{@{}c@{}}0-0\&0-1\\ Deve\end{tabular} & 1-0 & 1-1 & \begin{tabular}[c]{@{}c@{}}1-0\&1-1\\ User\end{tabular} \\
\toprule
Tencent  & 60  & 15  & 75(56.8\%)                                              & 6   & 51  & 57(43.2\%)                                              \\
TikTok          & 51  & 65  & 116(77.9\%)                                             & 4   & 29  & 33(22.1\%)                                              \\
DingTalk        & 371 & 247 & 618(74.7\%)                                             & 37  & 172 & 209(25.3\%)                                             \\
Keep            & 284 & 42  & 326(70.4\%)                                             & 28  & 109 & 137(29.6\%)                                             \\
Zoom            & 355 & 151 & 506(83.4\%)                                             & 44  & 57  & 101(16.6\%)                                            \\
\bottomrule
\end{tabular}
\label{tab6}
\end{center}
\end{table}

Answer to RQ3: This approach identified four feature feedback patterns (0-0), (0-1), (1-0), and (1-1) through the interaction between app update logs and user reviews. Combining these four patterns, it was observed that the level of user contribution varies among the five apps studied in their developmental evolution. Among them, Tencent Meeting had the highest user contribution, reaching 43.2\%, while Zoom had the lowest user contribution, at only 16.6\%. TikTok, DingTalk, and Keep had user contributions ranging from 22.1\% to 29.6\%.

\section{Related Work}
\label{sec:IV}
Analysis of app stores is a popular field of study in software engineering, and various methods have been developed in this area. In this section, we will discuss related work from two perspectives: 1) User review classification, and 2) Feature extraction and assessment. 

App stores serve as primary platforms for app developers to provide services and for users to share their opinions. Developers follow an iterative process when developing apps \cite{r20}, making it crucial to extract valuable information from user reviews, such as error reports and feature requests, which plays a vital role in the evolution of apps. However, user reviews lack predefined structural characteristics, often containing a significant amount of irrelevant information and expressing unstructured issues. Extracting meaningful insights and suggestions from user reviews is a hot topic in the field of software engineering. To address this challenge, much of the previous research has focused on classifying user reviews into different meaningful categories to filter out irrelevant reviews. Therefore, there has been significant work in review classification in previous studies \cite{r6, r13} to filter out irrelevant reviews. 

In previous work, categorizing the textual content of user reviews has been an important approach, but the approaches employed for grouping have a certain degree of subjectivity in research. User reviews can be categorized into multiple types when analyzed from different perspectives \cite{r5, r8, r10, r11, r12, r13}. Maalej et al. \cite{r8} classify user reviews into four basic categories based on the types of information contained in the review, including error reports, feature requests, user experience, and ratings. Guzman et al. \cite{r11}, expanding upon the research by Pagano et al. \cite{r5}, defined seven categories associated with app evolution. And given that a single user review might contain multiple categories \cite{r5}, Gu et al. \cite{r10} divided user reviews into sentences and identified five review types: aspect evaluation, error report, feature request, praise, and other. Panichella et al. \cite{r12} derived preliminary category definitions and descriptions through manual examination of developer mailing lists and then systematically mapped them to the classification proposed by Pagano et al. \cite{r5}, ultimately specifying four categories: information giving, information seeking, feature request, and issue discovery. Stanik et al. \cite{r13}, with the support of an innovation center in a large Italian telecommunications company, developed annotation guidelines to describe their understanding of problem reports, requests, and irrelevant reviews. Chen et al. \cite{r6} introduced AR-Miner, introduced AR-Miner, where user reviews were classified into informative and non-informative reviews, and non-informative reviews were subsequently filtered out.

However, even when reviews are classified for different reasons, app developers cannot respond to all user requests. They must prioritize and evaluate the importance of each request. Review analysis tools such as AR-Miner \cite{r6} and CLAP \cite{r21} perform clustering of user reviews and assign high or low priority to each review cluster, with high-priority clusters representing topics for developers to address in the next version. AR-Miner, for instance, filters out non-informative reviews and ranks informative reviews. In contrast, CLAP categorizes user reviews explicitly, and historical information for different categories is considered in priority ranking. Villarroel et al. \cite{r14} argued that issues mentioned by a larger number of users should be given priority, and they designed a clustering method to identify user reviews expressing similar opinions. Keertipati et al. \cite{r22} approached the issue from the perspective of software vendors, prioritizing user opinions extracted from review data. However, these works focus solely on text data generated by general users and do not address the issue of matching this data with app developer-side data. 

The findings of Wang et al. \cite{r23} suggest that app’s update logs partially reflect user requirements addressed in subsequent versions of the app. They observed that using app update logs to categorize user requirements in user reviews is not very helpful. In contrast to previous app-level analyses, Hassan et al. \cite{r24} conducted update-level research on user reviews. They combined app update information and user reviews to investigate detrimental updates, defined as updates that generate negative reviews from the general user base, to capture user sentiments towards specific updates. Gao et al. \cite{r25} introduced AOLDA, proposing the detection of issues in the current version based on statistical data from previous versions. AOLDA considers semantic relevance and user sentiment, evaluating each topic with the most relevant phrases and sentences, providing indications of issues that the app should prioritize.

Johann et al. \cite{r15} introduced SAFE, utilizing predefined text patterns to extract features from app pages and user reviews in the app store and matching them. Malik et al. \cite{r26} employed sentiment analysis to investigate user opinions on popular features, enabling feature recommendations for similar apps based on users preferences. Wu et al. \cite{r17} proposed KEFE, which identifies app key features through app description information and user reviews. Additionally, to understand common release strategies for mobile app, the underlying principles, and their perceived impacts on users, Nayebi et al. \cite{r27} conducted two surveys involving both users and developers. They identified differences between release strategies for mobile apps and traditional software, suggesting a need to reevaluate release decisions and adapt them based on user requirements and convenience.

\section{Discussion and Future Work}
\label{sec:V}
\begin{figure}[htbp]
\centerline{\includegraphics[scale=0.32]{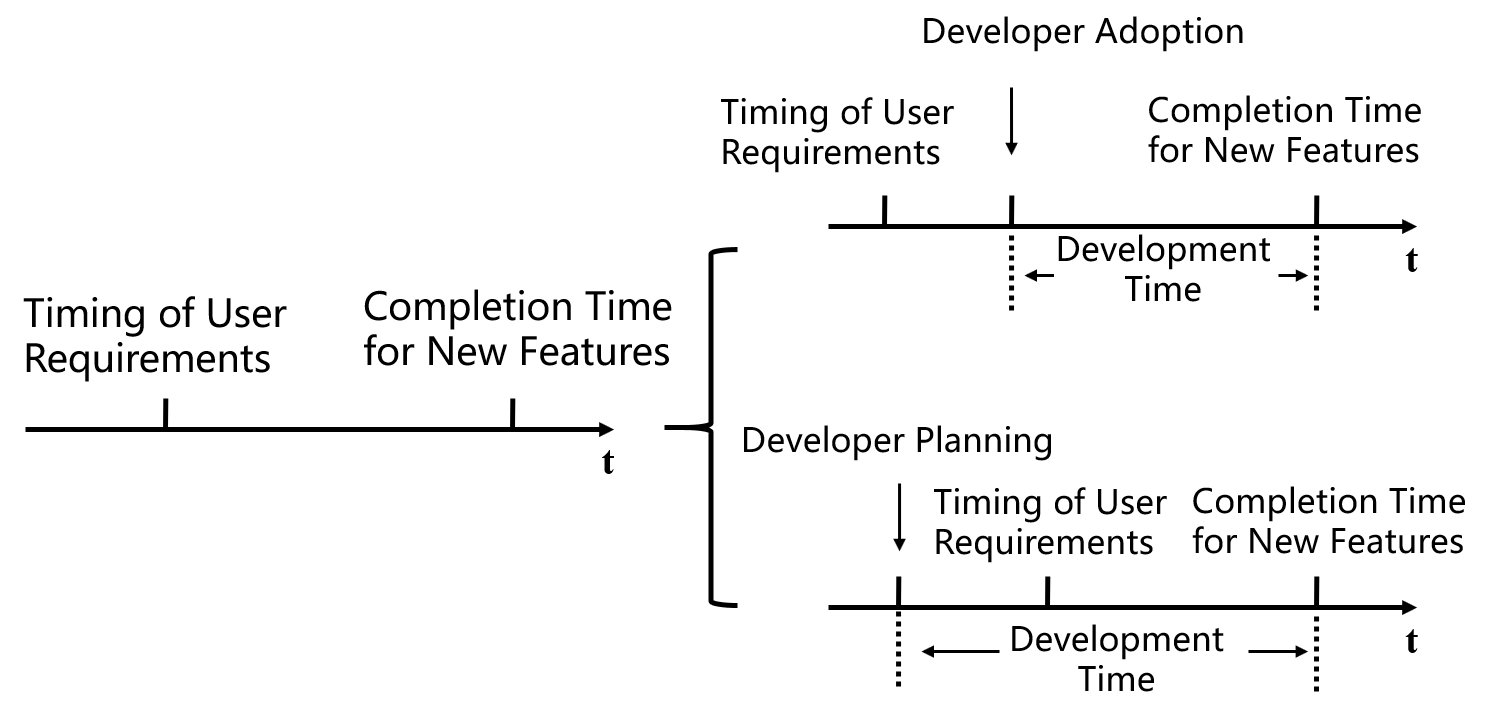}}
\caption{Developer Adoption and Planning Time}
\label{fig9}
\end{figure}

The work in this paper is to quantitatively analyze user contribution in app evolution based on data derived from user reviews and app update logs in the App Store. Specifically, our work focuses on establishing the matching relationship between user reviews and app update logs through text analysis and examining the temporal sequence of user review creation and app updates (i.e., release of update logs). Note that, the actual development of an app includes various stages such as programming and testing. Developers require a specific amount of time to transition from planning the development of new features to their actual implementation. In this paper, for any new feature in an update log, we use its release time as a dividing line.  By assessing whether the time when user requirements are raised precedes this time anchor point, we determine whether the release of the new feature is driven by user requirements, thus measuring user contribution. A potential scenario involves developers planning a feature, even before user suggestion. Because of constraints in development time, the release of this new feature may occur later than the time when users initially made their requirements, as shown in Figure~\ref{fig9}. For such cases, the new feature is not a result of user-initiated requirements. Instead, it is a feature that developers have planned and subsequently optimized and released after accepting user suggestions. The current approach proposed in this paper does not provide a detailed categorization for this specific scenario. We observe that the time cycles for releasing new versions vary among different apps. Even for the same app, identifying a regular pattern for developers' release times is challenging, making it difficult to determine developers' development time. Therefore, we currently do not incorporate the factor of development time into the approach presented in this paper.

In future work, to optimize the approach proposed in this paper and improve accuracy, we plan to engage with and consult app developers and vendors. This involves gaining insights into the actual development cycles of apps and determining the time developers require from planning new versions to releasing them. Building upon the approach proposed in this paper, we intend to introduce a buffer time for feature development response, considering the variable factors on the developer's side. This adjustment aims to provide a more precise measurement of new features being driven by user requirements. Furthermore, we will continue to acquire data from additional apps, delving deeper into the user contribution of different apps within the same category on the App Store. This will offer a more intuitive comparison for both developers and users, contributing to a more accurate and comprehensive analysis of user contributions and their impact on app evolution.

\section{Conclusion}
\label{sec:VI}
This paper introduces a time-aware approach to quantify user contributions in app evolution by mining the interactions between user reviews and app update logs. To effectively capture the interactions between users and apps, the proposed approach first employs a BERT classifier to classify user reviews into those related to app features and those unrelated. Subsequently, another BERT classifier is employed to match user reviews related to app feature with app update logs. Finally, through statistical analysis, the study explores feature feedback patterns to uncover user contributions. The approach is trained and validated on a dataset comprising 2,178 app update logs and 4,236,417 user reviews from five different apps. Experimental results demonstrate the effectiveness of the approach in both user review classification and matching, with average F1-Scores of 91\% each. Ultimately, the research reveals that user contributions vary among different apps in their developmental evolution. For instance, Tencent Meeting exhibits the highest user contribution rate at 43.2\%, while Zoom has the lowest contribution rate, at just 16.6\%.

\bibliographystyle{ieeetr}
\bibliography{ref}

\end{document}